\documentstyle[12pt]{article}
\newcommand{\be}{\begin{equation}}
\newcommand{\ee}{ \end{equation}}
\newcommand{\ba}{\begin{array}}
\newcommand{\ea}{\end{array}}
\newcommand{\NP}[3]{{\em Nucl. Phys.}{ \bf B#1#2#3}}
\newcommand{\PRD}[2]{{\em Phys. Rev.}{ \bf D#1#2}}

\newcommand{\IMP}[1]{{\em Int. J. Mod. Phys.}{ \bf A#1}}
\newcommand{\PL}[3]{{\em Phys. Lett.}{ \bf B#1#2#3}}

\begin{document}

\title{ \vspace{-15mm}
    {\normalsize \hfill
     \begin{tabbing}
    \` \begin{tabular}{l}
  HUB--EP--95/25 \\
  October 1995 \\
   hep--th/9510178\\
      \end{tabular}
      \end{tabbing} }
   \vspace{10mm}
   \setcounter{footnote}{1}
String-string duality for some black hole type solutions}
\author{~\\
\setcounter{footnote}{2}
Klaus Behrndt\thanks{e-mail: behrndt@qft2.physik.hu-berlin.de, Work
 supported by the DFG and a grant of the DAAD} \quad and \quad
\setcounter{footnote}{6}
   Harald Dorn\thanks{e-mail: dorn@qft3.physik.hu-berlin.de} \\
{\normalsize \em  Humboldt--Universit\"at, Institut f\"ur Physik } \\
   {\normalsize \em Invalidenstra\ss{}e 110, 10115 Berlin, Germany}
    }
   \vspace{10mm}
\date{~}
\maketitle

\begin{abstract} \noindent
We apply the duality transformation relating the heterotic to the IIA string
in 6D to the class of exact string solutions described by the chiral null
model and derive explicit formulas for all fields after reduction to 4D.
If the model is restricted to asymptotically flat black hole type solutions
with well defined mass and charges the purely
electric solutions on the heterotic side are mapped to dyonic ones on the IIA
side. The mass remains invariant. Before and after the duality transformation
the solutions belong to short $N=4$ SUSY multiplets and saturate the
corresponding Bogomol'nyi bounds.
\end{abstract}
\baselineskip=17pt
\renewcommand{\arraystretch}{2.0}
\thispagestyle{empty}
\newpage

In recent times it has been conjectured that different types of string
theory can be related by different types of duality transformations.
One example is the string--string duality \cite{du/kh,du/li/ra} relating
heterotic string theory compactified on $T^4$ and the type IIA string
theory compactified on $K3$ \cite{hu/to}.  Both theories live in $D=6$
and are related to each other by the field transformation

\be
\tilde{\Phi} = - \Phi \qquad , \qquad \tilde{G}_{\hat M\hat N} = e^{-2 \Phi}
G_{\hat M\hat N} \qquad , \qquad \tilde{H}_{\hat M\hat N\hat P}
= e^{-2 \Phi} ~^{*}H_{\hat M\hat N\hat P}~,
\label{1}
\ee
where $G_{\hat M\hat N}$, $\Phi$ and $H_{\hat M\hat N\hat P}$ are the metric,
dilaton and
torsion in the stringy frame. The canonical metric remains
invariant.

As one example it has been shown that the fundamental string solution
on the heterotic side becomes an instanton solution on the type IIA
side \cite{ha/st}. In this paper we are going to discuss a much more
general class of exact string backgrounds. In $D=6$ this class is
the chiral null model \cite{ho/ts, be}, which contains the fundamental string
and gravitational waves as special examples, and after the
reduction to $D=4$ it represents many known massive and massless
black hole solutions as well as Taub-NUT geometries.

First, we will discuss the heterotic side and recall some known facts
about this model. In a second step we are looking on the dual theory
and discuss the type IIA duality transform of known heterotic black/white
hole solutions.

\vspace{5mm}

On the heterotic side we start with the $D=6$ effective action
\be
S = \int d^6 x \sqrt{G} e^{-2 \Phi} \left[ R + 4 (\partial \Phi)^2 -
   \frac{1}{12} H^2 \right] \ .
\label{2}
\ee
We are assuming that the reduction from $D=10$ is trivial, i.e.\ we
have set all gauge fields and scalars coming from this reduction to zero
and will ignore furthermore this part of the internal space.
We are interested in the following solution of the equations of motion
\cite{ho/ts}
\be
\ba{l}
ds^2~ = ~ -2F(x) du [ dv - \frac{1}{2} K(x) du + \omega_I dx^I] ~ + ~ dx^I
       dx^I \\
B~=~ - 2F(x)du\wedge [dv+\omega_Idx^I] \quad , \quad e^{2\Phi}~=~ F(x)~,
\ea
\label{3}
\ee
with
\be
\partial ^2~F^{-1}~=~\partial ^2~K~=~0~,~~~~~\partial ^I \partial _{[I}~\omega
_{J]}~=~0~.
\label{3a}
\ee
In the following we call $v=x^0,~u=x^5$. Indices take the values:
latin from 1 to 3, greek from 0 to 3, large latin from 1 to 4 and hatted
large latin from 0 to 5.
This model is a generalization of the gravitational waves ($F=1$)
and the fundamental string ($K=1,0$\footnote{Vanishing $K$ is
possible here, but it is singular after reduction to $D=4$.}) background.
It possesses unbroken supersymmetries and is exact in the $\alpha'$
expansion.

Assuming that the fields depend only on $x^i$ and that $v=x^0$
becomes the time ($u$ and $x^4$ are the internal coordinates) we can
reduce this theory to $D=4$ (for details see Ref.\cite{se1}\footnote{We use
the conventions of \cite{se1} applied to the reduction from 6 to 4 dimensions.
We denote by $\vec{A}_{\mu}^{(1)}$
and $\vec{A}_{\mu}^{(2)}$ the internal space vectors
$(A_{\mu}^{(1)},A_{\mu}^{(2)})$ and $
(A_{\mu}^{(1+2)},A_{\mu}^{(2+2)})$, respectively.}).
Then the solution in $D=4$ in the canonical frame is given by
\be
\ba{l}
ds_{can}^2 = - e^{2\phi}(dx^0 + \omega_i dx^i)^2 + e^{-2 \phi}  dx^i dx^i~,
\\
\vec{A}^{(1)}_{\mu} = -\frac{1}{2}( F\omega_4 , 1)e^{4 \sigma} F\omega_{\mu}
\quad ,\quad
\vec{A}^{(2)}_{\mu} = \frac{1}{2}( \omega_4 , -K )e^{4 \sigma}
F^2\omega_{\mu}~,
 \\ e^{-4\sigma} = G_{44}G_{55}-G_{45}^2 = K F - F^2 \omega_4^2 \quad ,
\quad e^{-2\phi} = \sqrt{K F^{-1} - \omega_4^2}~,\\
h_{\mu \nu \rho} = 6e^{4\phi}\omega _{[\mu} \partial _{\nu}
\omega _{\rho ]}~,
\ea
\label{4}
\ee
with $\omega_{\mu} = ( 1 , \omega _i)$. The vector field
$\vec{A}^{(1)}_{\mu}$ comes from the metric and $\vec{A}^{(2)}_{\mu}$ from
the antisymmetric tensor, $h_{\mu \nu \rho}$ and $\phi $ denote the $4D$
antisymmetric tensor
and dilaton, respectively.
The moduli space of this model is parametrized by
the scalar fields of the theory which can be summarized to 3 complex fields
\cite{du/li/ra}
\be
\ba{l}
S = a + i e^{-2\phi}~,~~~~~~U = \frac{1}{T^*}~, \\
T = b + i e^{-2\sigma} = B_{45} + i \sqrt{G_{44}G_{55}-G_{45}^{2}} =
        F \omega_4 + i \sqrt{ K F - F^2 \omega_4^2}~,
\ea
\label{5}
\ee
where the axion $a$ is defined by
\be
h_{\mu \nu \lambda}~=~e^{2\phi}\sqrt{-g}~\epsilon _{\mu \nu \lambda \rho}~
g^{\rho \tau}\partial _{\tau}a~.
\label{5a}
\ee
Finally we can
combine the gauge fields to one graviphoton ($\vec{A}^{(+)}$) and two gauge
fields ($\vec{A}^{(-)}$) sitting in the vector multiplet by
\be
\ba{l}
\vec{A}^{(+)}_{\mu}
 = \frac{1}{\sqrt{2}} (\vec{A}^{(1)}_{\mu} + \vec{A}_{\mu}^{(2)})
 = -\frac{1}{2\sqrt{2}}  \,( \, 0 \, , \, 1 + |T|^2\, )\,e^{4 \sigma}
 F \omega_{\mu}~, \\
 \vec{A}^{(-)}_{\mu}
  = \frac{1}{\sqrt{2}} (\vec{A}^{(1)}_{\mu} - \vec{A}_{\mu}^{(2)})
  = -\frac{1}{2\sqrt{2}}\,( \, 2 b \, , \, 1 - |T|^2 \, )\,
  e^{4 \sigma} F \omega_{\mu} \ .
\ea
\label{6}
\ee
Before we go to the dual theory we summarize some properties of this
heterotic solution. First, we see that the solution is completely determined
by 4 harmonic functions: $F^{-1}$, $K$, $\omega_4$ and $a$ or equivalently
by the two complex scalars $S$ and $T$. Let us look on some special
examples.
We choose the following harmonic functions ($r^2= x^i x^i,~~R^2=x_1^2+
x_2^2+(x_3-i\alpha )^2$)
\be
\ba{l}
K= \mbox{Re}\left (1 + \frac{2m}{R}\right ),~~~
F^{-1} = \mbox{Re}\left (1 + \frac{2 \bar{m}}{R}\right),\\
\omega_4 =\mbox{Re}\left ( \frac{2q}{R}\right ),~~~a =
\mbox{Im}\left (\frac{2n}{R}\right ).
\ea
\label{7}
\ee
They describe a rotating black hole, see e.g. \cite{be} and refs. therein.
The corresponding mass, electric and magnetic charges
($g^{can}_{00}=-1+\frac{2M}{r},~~F_{0r}^{(\pm)}=\frac{\vec{Q}_{(\pm)}}{r^2},~~
^{*}F_{0r}^{(\pm)}=\frac{\vec{P}_{(\pm)}}{r^2}$ for large $r$) are given by
\be
\ba{l}
M = \frac{1}{2}(m + \bar{m}) \quad ,\quad
\vec{Q}_{(+)} = \frac{1}{\sqrt{2}} (\, 0 \, , \, 2 M \, ) \quad , \quad
\vec{Q}_{(-)} = \frac{1}{\sqrt{2}} (\,- 2q \, ,\, m-\bar{m}  \, )~.
\ea
\label{8}
\ee
Note there are no magnetic  charges.
The Bogomol'nyi bound contains only the $(+)$ sector and is saturated
$|Q_{(+)}|^2 = 2 M^2$.
Three cases are of special interest \cite{be,du/ra}
\newline
{\em 1.) massless black/white holes, $\bar{m} = - m$}
\newline
In this case the graviphoton sector is uncharged. These configuration have
an additional naked singularity at $r_0^2=2|Q^{(-)}|^2$ which is repulsive
\cite{ka/li} and cannot be obtained as an extremal limit of known black hole
solutions. They can be identified as string states with $N_R=\frac{1}{2}$,
$N_L=0$.
\newline
{\em 2.) Kaluza-Klein black holes, $\bar{m}m=0$}
\newline
These solutions are characterized either by a constant $F$ or constant
$K$, i.e.\ they correspond to the fundamental string or gravitational
waves in original $D=6$ theory. The string states are $N_R=\frac{1}{2}$,
$N_L=1$.
\newline
{\em 3.) extremal dilaton black holes, flat internal
space}
\newline
The flat internal space corresponds to $b=0$ and  $|T|=1$ ($q=0$,
$\bar{m}=m$). In this case the $T$ modulus is trivial
and we have only the graviphoton, i.e.\ in this case we have pure
gravity. The string states are given here by a discrete set with
$N_R=\frac{1}{2}$, $N_L>1$.

Of course, these are only some cases and the general solution
contains many others, but these solutions found special interest
in the literature. At the end we will return to these cases
and discuss their dual partners (type IIA).

\vspace{5mm}

In the next step, we perform the string/string duality in $D=6$
and reduce the dual model to $D=4$.
The duality transformation as defined in (\ref {1}) leads to
\be
\ba{l}
d\tilde{s}^2~=~-2du[dv-\frac{1}{2}Kdu+\omega _Idx ^I]~+~F^{-1}dx^Idx^I~,\\
e^{2\tilde{\Phi}}=
F^{-1}
\ea
\label{9}
\ee
immediately. A little bit more effort is needed to find the dual $\tilde B,~
\tilde H$. We start with
\be
H_{\hat M \hat N \hat P}~=~3\partial_{[\hat M}B_{\hat N \hat P]}
\label{10}
\ee
and find
\be
\tilde H _{ABC}=-\epsilon^{ABCI}\partial _I F^{-1}~,~~~~~
\tilde H _{AB5}=-\epsilon^{ABIJ}\partial _I \omega _J~.
\label{11}
\ee
All other components are zero.

As a side remark we would like to emphasize that the special case $K=\omega _I
=0$ just gives the axionic soliton background
\cite{ca/ha/st}. Hence, we get the known statement that the
elementary (or fundamental) string background on the heterotic side is mapped
to a soliton on the type II side \cite{ha/st} \footnote{For the inclusion
of an additional $v$-dependence see \cite{da}.}
\be
d\tilde{s}^2 = -2 du dv + e^{2\tilde{\Phi}} dx^I dx^I \quad , \quad
\tilde{H}_{ABC} = -\epsilon ^{ABCI} \partial _I e^{2\tilde{\Phi}}~.
\label{11a}
\ee
Our background is then a natural generalization of this dual pair. We should
also mention that $\omega _I=0,~K\neq 0,~F\neq 0$ has been generalized to
chiral
null models with curved transverse space \cite{cv/ts}.

To write down $\tilde B$ corresponding to (\ref{11}) we introduce
the following transformations (assuming
$\partial _4 =0$)
\be
\ba{l}
\partial _a \omega _4~=~\epsilon ^{abc}\partial _b \tilde{\omega}_c~,~~~~~
\partial _{[a}\omega _{b]}~=~\frac{1}{2} \epsilon ^{abc} \partial _c \tilde {
\omega}_4~,\\
\partial _a F^{-1}~=~ - \epsilon ^{abc}\partial _b \bar {\omega}_c~. \\
\ea
\label{12}
\ee
This system of differential equations for the new quantities $\tilde{\omega}
_I,~~ \bar{\omega}_i$ is soluble on shell. Note that via (\ref{4}),(\ref{5a})
$a=-\tilde{\omega}_4$
has been introduced as the axion already.
Using these new functions we get for the nonzero components finally
\be
\tilde B _{m4}~=~-\bar{\omega}_m,~~~\tilde B _{m5}~=~-\tilde{\omega}_m,~~~
\tilde B _{45}~=~-\tilde{\omega}_4.
\label{13}
\ee
Now we repeat the dimensional reduction on the dual side and get
\be
\ba{l}
d\tilde s ^2_{can}~=~ds _{can}^2,~~~\tilde{\phi}~=~\sigma
,~~~\tilde{\sigma}~=~\phi,\\
\tilde {\vec A}^{(1)}_{\mu} = -\frac{1}{2}( \omega_4 , F^{-1})
\omega_{\mu}e^{4 \tilde{\sigma}}~=~\vec A^{(1)}_{\mu},\\
\tilde {\vec A}^{(2)}_{\mu} = \frac{1}{2}(\bar{\omega}_{\mu},\tilde{\omega}
_{\mu})~-~\frac{1}{2}
(F^{-1},-\omega _4)\tilde{\omega}_4\omega _{\mu} e^{4 \tilde{\sigma}}.
\ea
\label{14}
\ee
To unify the notation we have introduced $\bar{\omega}_0=\tilde{\omega}_0=0$.
In contrast to the original case the 4D antisymmetric tensor turns out to be
nonzero
\be
\tilde b _{\mu \nu}~=~\omega _4 F^2 e^{4\tilde {\phi}}\omega _{[
\mu}\bar {\omega}_{\nu ]}~+~F e^{4\tilde {\phi}}\omega _{[\mu
}\tilde{\omega}_{\nu ]}.
\label{15}
\ee
Taking into account also the gauge field contribution \cite{se1} the related
field strength $\tilde h$ is
\be
\ba{l}
\tilde h _{mnl}~=~-e^{4\tilde{\phi}}\epsilon^{mnl}\omega _k \partial _k(F
\omega _4), \\
\tilde h _{0nl}~=~-e^{4\tilde{\phi}}\epsilon^{nlk} \partial _k(F
\omega _4).
\ea
\label{16}
\ee
\mbox{From} these equations we can read off the dual axion
\be
\tilde a~=~ F\omega _4~=~b.
\label{17}
\ee
Thus, as expected the $S$ and $T$ moduli interchange under string-string
duality
\be
\tilde S~=~T,~~~\tilde T~=~S,~~~\tilde U~=~U.
\label{18}
\ee
For the special type of solutions introduced in (\ref{7}) we find in spheroidal
coordinates $r,~\theta ,~\phi$
\be
\ba{l}
\omega_{\phi}~=~-\frac{2nr\alpha \sin ^2 \theta}{r^2+\alpha ^2
\cos ^2 \theta}~,\\
\tilde{\omega}_{\phi}~=~\frac{2q(r^2+\alpha ^2)\cos \theta}{r^2+\alpha ^2
\cos ^2 \theta}~,~~~
\bar {\omega}_{\phi}~=~ - \frac{2\bar m(r^2 + \alpha ^2)\cos \theta}
{r^2+\alpha ^2
\cos ^2 \theta}~.
\ea
\label{19}
\ee
Studying the asymptotic behaviour for large $r$ we can read off the mass,
electric and $magnetic$ charges
\be
\ba{l}
\tilde M~=~M~,\\
\tilde {\vec Q}_{(\pm)}~=~\frac{1}{\sqrt{2}}(-q,m)~,~~~~
\tilde {\vec P}_{(\pm)}~=~\pm \frac{1}{\sqrt{2}}
(-\bar m, q).
\ea
\label{20}
\ee
\mbox{From} the technical point of view the appearance of magnetic charges in
the dual
case is due to the $O(1)$ asymptotic behaviour of $\tilde{\omega}_{\phi}$
and $\bar {\omega}_{\phi}$ compared to the fall off for $\omega_{\phi}$.
We expect the values of $q,~m,~\bar m$ to be restricted by the dyon
quantization condition in such a way that $q^2\pm m\bar m,~~2qm,~~2q\bar m$
are integers.

Furthermore, since now electric and magnetic charges are present we have
to consider the Bogomol'nyi bound in its generalized forms
\cite{ka, cv/yo, du/li/ra}. For the type IIA case both sectors $(\pm )$ are
involved \cite{du/li/ra}
\be
\ba{l}
\tilde{M}^2~\geq \mbox{Max}\left ( \vert \tilde{Z}_1 \vert ^2~,~ \vert
\tilde{Z}_2 \vert ^2 \right )
{}~,\\
\vert \tilde{Z}_{1/2}\vert ^2~=~\frac{1}{2}\left [ (\tilde{Q}^{(\pm)}_1\pm
\tilde{P}^
{(\pm)}_2)^2~+~(\tilde{Q}^{(\pm)}_2\mp \tilde{P}^{(\pm)}_1)^2 \right ]~.
\ea
\label{21}
\ee
$Z_1$ and $Z_2$ denote the two central
charges in the $N=4$ SUSY algebra. With (\ref{20}) we get
\be
\vert \tilde{Z}_1 \vert ^2~=~ \vert \tilde{Z}_2 \vert ^2 ~=~\tilde{M}^2~.
\label{22}
\ee
Like the original heterotic solution our transformed IIA solution belongs to a
short multiplet. Again the Bogomoln'yi bound is saturated. Concerning the
identification of string states as usual only pure
electric solutions can be related to elementary string states. Nonvanishing
magnetic charges have to be attributed to solitonic states. In our case
$\tilde{\vec P}=0$ implies Kaluza-Klein black holes with
$\tilde M=\frac{1}{2} m,~~\tilde {\vec Q}=
(0,\frac{1}{2}m)$. According to the type IIA mass formula this means
$\tilde{N}_L = \tilde{N}_R=\frac{1}{2}$ \cite{du/li/ra}. With respect to the
three special cases discussed on the heterotic side we observe that both the
transforms of
the massless case as well as the dilatonic black hole have magnetic charge
vectors orthogonal to the electric ones.\\[2mm]
{\bf Acknowledgement}\\
We would like to thank R. Kallosh, D. L\"ust and H.-J. Otto for useful
discussions.

\renewcommand{\arraystretch}{1.0}

\end{document}